\documentstyle[aps,twocolumn,epsf]{revtex}

\newcommand{\be}{\begin{eqnarray}}
\newcommand{\ee}{\end{eqnarray}}

\newcommand{\no}{\nonumber}

\renewcommand{\vec}{\bbox}

\begin{document}

\title{Phase diagram of 3D  $SU(3)$ gauge-adjoint Higgs system }

\author{S. Bronoff, R. Buffa and C. P. Korthals Altes}

\address{Centre Physique Th\'eorique au CNRS, Case 907, Luminy,
F13288, Marseille, France}

\date{\today}

\maketitle

\begin{abstract}
Thermally reduced QCD leads to three dimensional $SU(3)$ gaugefields
coupled to an adjoint scalar field $A_0$. We compute the effective potential
in the one-loop approximation and evaluate the VEV's of $TrA_0^2$ and $TrA_0^3$.
In the Higgs phase not only the former, but also the latter has a VEV.
This happens where the $SU(3)$ gauge symmetry is broken minimally with U(2) still unbroken. It is plausible  that in the Higgs phase one has a transition for large enough Higgs selfcoupling to a region where $TrA_0^3$ has no VEV and where the gaugesymmetry is broken maximally to $U(1)\times U(1)$. 
\end{abstract}


\narrowtext


\section{Introduction}
\label{sec1}


To obtain reliable information about the quark-gluon plasma well above the
critical temperature thermally reduced QCD is extremely efficient.
  
Since its inception~\cite{dimred} it has been pioneered in recent years~\cite{debye} for a precise evaluation of the Debye mass for $SU(2)$ and $SU(3)$ gauge groups with or without quarks.

In this note we want to point out some properties of the phase diagram
of $SU(3)$. These properties have to do with the Higgs phase of the three
dimensional gauge-adjoint scalar theory. They are qualitatively different from 
its $SU(2)$ counterpart. 

The main point of this paper is  how and what breaking patterns of the various symmetries in the Lagrangian do emerge. R-symmetry breaking and gauge symmetry
breaking are narrowly related, and we surmise where they are realized in the phase diagram.  
The Higgs phase in our reduced action is induced by quantum effects.
These effects are calculable for certain values of the parameters 
in the Lagrangian by a loop expansion~\cite{loop}. This calculation
results in a critical line in the phase diagram. Below this line 
the Higgs phase realizes, with the VEV of $TrA_0^2$ non-zero. But in 
$SU(3)$ the VEV of $TrA_0^3$  is not necessarily zero, and it acquires a VEV
in the loop expansion. At the same time  the gauge symmetry is minimally
broken, leaving a $U(2)$ group unbroken. This phase is obviously absent in $SU(2)$.
However a phase where simultaneously R-parity is restored and the gauge symmetry is maximally broken to $U(1)\times U(1)$ is possible and we surmise it is realized
in between the symmetric phase and the broken R-parity phase discussed above. This would be a phase in which Abelian monopoles screen the two photons
just as in the SU(2) case~\cite{polyakov}. Some aspects of the R-parity 
breaking have been briefly mentioned LAT98~\cite{altes}.

\section{Reduced QCD Lagrangian} 
\label{sec2}


The Lagrangian of reduced QCD reads:
\be
S=\int d\vec x&\{&{1\over 2}\sum_{i,j}TrF_{ij}^2+\sum_iTr(D_iA_0)^2+
  m^2TrA_0^2\no\\
&+&\lambda(TrA_0^2)^2+\delta S\}
\label{redaction}
\ee

The term $\delta S$ stands for all those interactions consistent with gauge,
rotational and R-symmetry (i.e. $A_0\to -A_0$). These are  symmetries
respected by the reduction.

There is a remarkable cancellation, to two loop order, for the coefficients
of all renormalizable and nonrenormalizable terms containing only $A_0$~\cite{these}.

We will study the reduced  action limited to the superrenormalizable 
terms in eq.~\ref{redaction}. The gauge coupling $g_3^2$ has dimension of
mass, so the phase diagram is specified by two dimensionless parameters~\cite{loop} $x=\lambda/g_3^2$ and $y=m^2/g_3^4$. When related to the underlying 4D physics, the temperature T and the gauge coupling g, it is obvious that $x\sim g^2$and $xy\sim constant$ at very high temperature.  Hence the choice of axes in the phase diagram.

\section{The effective action}
\label{sec3}

Understanding the phase diagram of the 3D theory necessitates the 
knowledge of the effective action. We define it as:
\be
\label{effaction}
\exp{-VS_{eff}(C,E)}=\int DA&\delta&\left(g_3^2C-\overline{TrA_0^2}\right)\cdots\no\\
\cdots&\delta&\left(g_3^3E-\overline{TrA_0^3}\right)\exp{-S}
\ee

The bar means the  average over the volume V. Hence the volume factor
in front of the effective action on the l.h.s. in eq.~\ref{effaction}.
In the large volume limit we can deduce the effective action for $TrA_0^2$
by minimizing over E, keeping C fixed:

\be
\label{efffaction1}
min_{\{E\}}S_{eff}(C,E)=V(C)
\ee

The effective action for $TrA_0^3$ follows similarly:
\be
\label{efffaction}
min_{\{C\}}S_{eff}(C,E)=W(E)
\ee

Important is to grasp the physical meaning of $V$ and $W$.  $\exp{-VV(C)}$ is proportional to the probability that we measure
the value C for the averaged order parameter $\bar TrA_0^2$ in the system
enclosed in a volume V. It is therefore a gauge invariant quantity.
Similar for W(E). It should  not be confused with the 1PI effective action. The reader can find more detail on this point in ref.~\cite{these}.
It is very useful to express the effective action $S_{eff}(C,E)$ is  in terms of the variables $B_1, B_2$ and $B_3$, the diagonal elements of a traceless diagonal
$3\times 3$ matrix $B$. This matrix $B$ is the background field of $A_0$, and so, apart from from
fluctuations:
\be
g_3^2C=TrB^2;\hspace{4mm} g_3^3E=TrB^3
\ee
The matrix B can be expressed in the two diagonal generators of SU(3),
normalised at $1/2$:
{\small\be
B&=&q (2\sqrt3) \lambda_8+2r\lambda_3=\mathrm{diag}(q/3+r/2,q/3-r/2,-q/3)
\ee}

In fig. 1 we have plotted the symmetries of the effective action in this plane,
due to permutation symmetry of the $B_i$ and R-symmetry $B\to -B$.
At the same time we show the curves of constant E, and in particular the
lines where E=0. These are the directions where the gauge symmetry breaking
is maximal, e.g. $\lambda_3\sim diag(1/2,-1/2,0)$. So in a phase with $TrA_0^2\neq 0$ the statement that $TrA_0^3$
has no VEV means the symmetry is broken maximally and vice versa.

\begin{figure}\epsfxsize=6cm
\centerline{\epsfbox{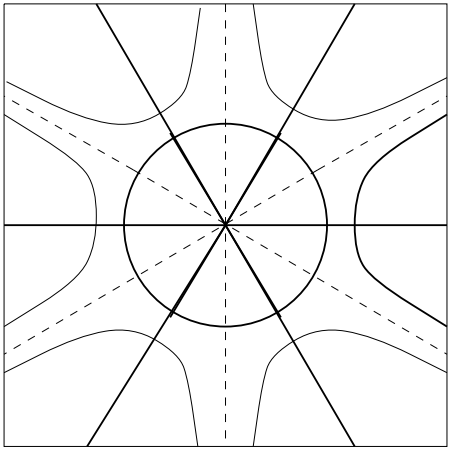}}
\vskip4mm
\caption{The horizontal axis is the $\lambda_8$, vertical direction the $\lambda_3$ direction. Permutation and R-symmetry render the potential identical
in the six segments between either broken or continuous lines.
The former are directions of maximal, the latter of minimal gauge 
symmetry breaking. The circle is defined by fixing C, the cubic invariant
fixes the curves with the broken lines as asymptotes.}
\label{fig1}
\end{figure}

Is there an analogous statement for the phase where the cubic invariant 
has a non-zero VEV? From the fig. 1 it is clear that curves of constant
$TrA_0^3\neq 0$ correspond to a minimum value of $TrA_0^2$ where
the breaking is in the $\lambda_8$ direction ($\sim diag (1,1,-2)$ or one of its permutations) in which directions  the gauge symmetry 
breaking is minimal. In section IV we show that a minimum value of $TrA_0^2$
indeed corresponds to a minimum value of the potential at constant E.

So the cubic invariant is non-zero if and only if the gauge symmetry breaking is minimal, i.e. U(2) is still unbroken. 

\section{Which phases are realized?}
\label{sec4}


To answer this question we try the loop expansion.
The loop expansion will be only trustworthy where we
find a Higgs phase, hence masses for our propagators.
In the symmetric phase Linde's argument~\cite{linde} will apply.

The loop expansion starts from the background field B and admits
for small fluctuations around B:
\be
A_0&=&B+Q_0\no\\
A_i&=&Q_i
\ee

One does do a saddle point approximation around $B$ in the path integral in  eq.~\ref{effaction}.
The gauge invariant constraints have to be taken into account.
Most convenient is to introduce two variables $\gamma$ and $\epsilon$
to Fourier analyse the deltafunction constraints in ~\ref{effaction}.
Like for the fields, we split them into background and quantum variables:
$\gamma=\gamma_c+\gamma_{qu}$ and like wise for $\epsilon$.
One does do a saddle point approximation around $B$, $\gamma_c$ and $\epsilon_c$
 in the path integral in  eq.~\ref{effaction}.
The linear terms in $\gamma_{qu}$, $\epsilon_{qu}$, $TrB\bar Q_0$, and 
$TrB^2\bar Q_0$ give respectively equations of motion:
 
\be
g_3^2C-TrB^2=0\\
g_3^2E-TrB^3=0\\
-2i\gamma_c +V\left(2m^2+4\lambda TrB^2\right)=0\\
\epsilon_c=0
\label{motion}
\ee

The quadratic part in the expansion of $Q_0$ contains a term from the constraint:
\be
-i\gamma_c\bar TrQ_0^2+3i\epsilon\bar TrBQ_0^2 
\ee
apart from the usual contribution from the reduced action, eq.~\ref{redaction}.
 
Substituting from the equations of motion~\ref{motion} into the quadratic
 part eliminates all terms proportional to $Tr Q_0^2$ and $TrBQ_0^2$.
Only terms
proportional to $\lambda (TrBQ_0)^2$ stay. So all Higgs masses are zero,
except for the diagonal component $TrBQ_0$\footnote{This is in contrast
with methods using the 1PI functional.} . The only way they still can get masses is through the gauge fixing. This is
 essential for the 
gauge independence as we shall see now.
 
By choosing $R_{\xi}$ gauge we eliminate mixed terms between $Q_0$ and $\vec Q$
. The only $\xi$ dependence enters in the masses of the off-diagonal ghosts, 
longitudinal vectorbosons and Higgs. For a given off-diagonal component (ij), ($1\le i\le j\le N$) they  all equal $\xi g_3^2(B_i-B_j)^2$. Thus integration over these modes gives cancellation of the gauge dependent masses.
The diagonal Higgs do not get a mass from  $R_{\xi}$ gauge and the result 
to one loop order reads:
\be
S_{eff}(C,E)/g_3^6&=&yC^2+xC^4\no\\
&-&{1\over{3\pi}}\{\sum_{i\le j}\vert B_i-B_j\vert^3+2(xTrB^2)^{3/2}\}
\ee
This is valid for number of colours $N=2,3$.
The correction $O(x^{3/2})$ in this formula is actually smaller than the
correction of order $O(x)$ from the two loop contribution.
\begin{figure}\epsfxsize=6cm
\centerline{\epsfbox{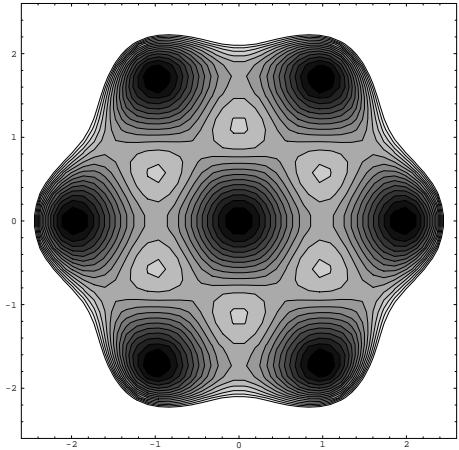}}
\vskip4mm
\caption{The SU(3) potential, eq.(8), on the critical line
as explained in the text, at x=0.1. Horizontal and
vertical axis as in fig. 1.
Dark areas show low, light ares high values. }
\label{fig2}
\end{figure}

As noted by the authors of ref.~\cite{loop} the minus sign in front of the induced cubic term is reminiscent of  an invariance present in 4D : $Z(N)$ invariance. The latter is broken by the presence of the term $(TrB^2)^{3/2}$, which for x small can be neglected. 

The resulting surface tension $\alpha$ is for small $x\sim g^2$
\be
\alpha\sim{1\over g}(1+O(g^2))\nonumber
\ee
and drops when moving to the right on the critical line. This is as expected from the SU(2) case\cite{loop}.

Our result for $N=2$ differs slightly from that quoted in ref.~\cite{loop} eq. 3.4. We find the loop expansion does introduce only $x$-dependence,
no $y$ dependence. This for two simple reasons: $y$ obviously never appears in vertices, and in the quadratic part it drops out through the equations of motion for the saddle point. So it does not appear in propagators either.
 The result is gauge choice independent, as it should.

In fig. 2 we have a contour plot of the potential for N=3 on the curve 
\be
xy_c={3\over{8\pi^2}}+O(x)
\ee
This is the locus in the (x,y) plane of degenerate minima along the $\lambda_8$ direction and its equivalents.

To obtain the potentials (3) and (4)
we have a simple analytic proof of where the minima are, but  
the reader can see by inspection that both the effective potentials for 
the quadratic and for the cubic invariants are determined by the minimum
along the $\lambda_8$ direction, using eq.~\ref{efffaction} and the curves
of constant C and constant E in fig.1.

Hence where the loop expansion is valid the phase with minimal gauge symmetry
breaking  and R-symmetry breaking realizes. Note that in this direction
the VEV's are related by:

\be
\vert TrA_0^3\vert={1\over \sqrt 6} (TrA_0^2)^{3/2}
\ee

\begin{figure}\epsfxsize=7.3cm
\centerline{\epsfbox{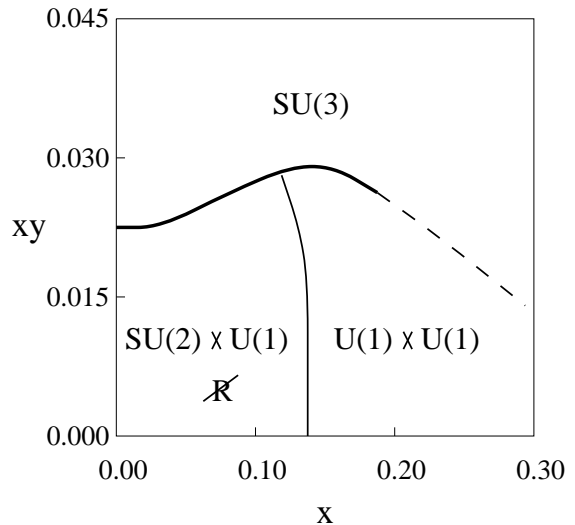}}
\vskip4mm
\caption{Sketch of the SU(3) phase diagram, with unbroken
gaugegroups indicated. R-symmetry is only broken in the left
lower phase}
\label{fig3}
\end{figure}

\section{A second Higgs phase?}
\label{sec5}

Having established the presence of a phase with minimal symmetry
breaking for small x, we have to look for approximation methods 
at large x. One possibility is mean field with a loop expansion
at large values of the lattice coupling and we are analyzing this~\cite{these}.
Also under study is a lattice Monte-Carlo study~\cite{boucaud}.

As we said earlier, this is the phase where Polyakov's mechanism is
at work~\cite{polyakov}. The two photons are screened by the Abelian
fields of the 't Hooft Polyakov monopoles~\cite{thooft}. This phase
would in part of the phase diagram (see fig. 3) join smoothly the symmetric
phase, like in SU(2).
\section{Summary}
\label{sec6}

In summary, we have shown that at least one Higgs phase is realized with spontaneously
broken R-symmetry and minimal gauge symmetry breaking. It is very plausible that the other phase is located as shown in fig. 3. Montecarlo simulations
should test for the additional Higgs phase. 

Very interesting is the contrast between the SU(2) and SU(3) monopole
case. What about the non-Abelian monopoles
in the minimally broken phase in SU(3)? Monte-Carlo studies  as in ref.\cite{teper} should reveal
their structure. Some analytic progress on classification and dynamics
of these monopoles has been made by K. Lee and Bais~\cite{bais}

\acknowledgments

The authors thank the ENS for its kind hospitality when this work was
done and Philippe Boucaud for interest and incisive remarks. The referee's remarks were useful for clearing up some of the issues. C.P.K.A. thanks Sander
Bais and Jan Smit for useful discussions. Both S.B. and R. B. are indebted
to to the MENESR for financial support.

\end{document}